\documentstyle[pre,epsfig,aps]{revtex}
\def\ee{\end{equation}}
\def\be{\begin{equation}}
\def\ba{\begin{eqnarray}}
\def\ea{\end{eqnarray}}
\begin{document}
\twocolumn[
\hsize\textwidth\columnwidth\hsize\csname@twocolumnfalse\endcsname
%
\title{Zeroth law of thermodynamics and the transformation \\ from  
nonextensive to extensive framework}
\author{Ramandeep S. Johal}
\address{Institute for Theoretical Physics,
Technical University, D-01062 Dresden, Germany }
\maketitle
\begin{abstract}
Within the nonextensive framework, it is shown that zeroth law of 
thermodynamics determines not only the mapping between
Lagrange multipliers and intensive variables,
but also the mapping between nonextensive and extensive
entropy. The form of constraints decides
the form of the extensive entropy, standard averages
lead to Boltzmann-Shannon-Gibbs entropy while
normalised biased averages lead to Renyi entropy.
The mapping between Lagrange multipliers and intensive variables
is also discussed in the more general context of composable entropy.  

\noindent {\it PACS Number(s):  05.20.-y, 05.70.-a, 05.90.+m}
\end{abstract}
]
%
In recent years, there has been a great deal of interest in the
generalized entropy satisfying the following non-additive
property
\be
S_q(A+B) = S_q(A) + S_q(B) + (1-q)S_q(A)S_q(B),
\label{nad}
\ee
where $A$ and $B$ represent two independent
subsystems composing  a larger system $(A+B)$ and
$q$ is a real parameter. This form of non-additivity
 is the simplest example \cite{abepre01} for a composable entropy
consistent with thermodynamic equilibrium.
Let $p_i$ be the probability distribution set characterising 
the discrete microstates $i=1,...,W$ of the system. 
 The generalized entropy satisfies the condition that for $q\to 1$, $S_q$
goes to Boltzmann-Gibbs-Shannon (BGS) entropy,
$S=-\sum_{i=1}^W p_i \ln p_i$ and one recovers the usual
additivity of entropy from (\ref{nad}).
An explicit realisation of such an entropic form is
popularly known as  Tsallis entropy \cite{tsallis88}
\be
S^{(T)}_q = \frac{1-\sum_{i=1}^W p_i^q}{q-1}.
\label{st}\ee 
It is concave for all positive values of $q$ and  has
been proposed as a basis to generalize the  
standard Boltzmann-Gibbs (BG) thermostatistics.
The motivation for this proposal is to  be able to treat the
nonextensivity inherent in many physical systems
and also to deal with those anomalous systems where the
BG formalism fails to give well defined results \cite{tsallisspver}.
 Indeed, nonextensivity plays important role in systems with
long range interactions  such as astrophysical systems \cite{nexastro},
 ferroic materials \cite{wadhawanbook}, 
systems with non-separable or overlapping parts such as in 
quantum entanglement \cite{canosaprl02}, and broadly speaking in 
many nonequilibrium phenomena.

In the Tsallis formalism, a maximisation problem for $S^{(T)}_q$
is formulated, imposing  as constraints the generalized
 mean values of the form \cite{tmp98} 
\be
{\sum_{i=1}^{W}  p_{i}^{q} E_{k,i} \over \sum_{j=1}^{W} p_{j}^{q} } 
= E_{k}^{(q)},
\qquad (k=1,...,n),
\label{cons1}
\ee
alongwith the normalisation of probability, $\sum_{i=1}^{W}p_{i} =1$.
Thus it may be taken as a generalization of Jaynes' approach
to statistical mechanics.
The resulting  non-canonical distributions are found to
be  power-law type,
\be
 p_i\sim \left[1 - (1-q)\sum_{k=1}^{n} \eta_{k}^{(\rm T)} (E_{k,i}- E_{k}^{(q)})/
\sum_j p_{j}^{q} \right]^{1/(1-q)},
\label{eqpd}
\ee
where $\eta_{k}^{(\rm T)}$ are the $n$ number of Lagrange multipliers
associated with the constraints (\ref{cons1}).
The Tsallis statistical weight 
reduces to the usual exponential Boltzmann factor, 
 for $q\to 1$.  Such generalised distributions have
been found to fit the experimental data very well in many
different situations; some of the more recent applications 
are  fully developed turbulence \cite{nexturb}, financial markets \cite{fmichael},
anomalous diffusion in Hydra cellular aggregates \cite{upadhya01},
CMR manganites \cite{reis02} and hadronic jets \cite{bediaga00}, 
to name a few \cite{httpcbpf}.
Realising the apparent success of Tsallis-type distributions in
realistic situations, there has been a growing
interest to justify or clarify the premises of thermodynamic
formalism based on  the modified postulates.  
In this context, some of the important issues are:
 uniqueness of the generalised entropy \cite{santos97}, physical interpretation
of the nonextensivity index $q$ \cite{johal98} and possibility
of fixing its appropriate value from the theory \cite{lyrprl98}, and
statistical mechanical foundations for nonexponential distributions
\cite{rajaabedenton}.

Another  crucial issue  within the Tsallis formalism,
 has been   the interpretation of Lagrange 
multipliers \cite{rama00,abemartinez01} entering the maximum entropy problem.
It has been found that these parameters are not intensive
in nature, unlike their counterparts appearing in the standard
Boltzmann-Gibbs (BG) formalism and so are not the ones which control
mutual equilibrium between independent subsystems (such as
common temperature, pressure, etc.). Speaking in the context of 
canonical ensemble,  the appropriate
physical temperature has to be defined from the generalized
zeroth law of thermodynamics. 
Somewhat unexpectedly, the forms of  thermodynamic
relations are the same as in the extensive framework,
 when the calculations are done using the intensive 
variables \cite{abemartinez01,toralpp}.
More precisely, it has been suggested that  
a mapping from the nonextensive Tsallis formalism 
to an extensive thermodynamic formalism can be defined \cite{vives02}, 
which basically involves the following two transformations:
\be
{S^*}^{(e)} = { {\rm ln}\; [1 + (1-q){S^*}^{(n)}] \over (1-q)},
\label{ne}
\ee
where for a given {\it maximum} (denoted by $*$) 
nonextensive entropy ${S^*}^{(n)}$, 
there exists  an extensive entropy ${S^*}^{(e)}$  
with the same concavity as ${S^*}^{(n)}$, provided  $q<1$. Also, we have   
\be
\eta_k = {\tilde{\eta}_k \over [1 + (1-q){S^*}^{(n)}]  }.
\label{lagin}
\ee
Here $\tilde{\eta}_k$ is the Lagrange multiplier connected with the
maximisation of ${S^*}^{(n)}$ and $\eta_k$ is the corresponding intensive  
variable satisfying $\eta_k (A) = \eta_k (B)$ for two systems $A$ and $B$
in mutual thermodynamic equilibrium.
The transformations (\ref{ne}) and (\ref{lagin}) were obtained in \cite{vives02} by 
using scaling property of the nonextensive entropy
(\ref{nad}) for equilibrium states and requiring that 
the standard form of the Gibbs-Duhem equation 
is preserved. 
Now since this treatment 
assumes a priori,  the explicit form of nonextensive
entropy to be Tsallis type (\ref{st}), the extensive entropy
obtained from (\ref{ne}) is fixed to be Renyi entropy
of order $q$ \cite{renyibook}
\be
S^{(R)}_q = \frac{\ln \sum_i p_i^q}{1-q}.
\label{sr}
\ee
However, as the standard extensive thermodynamic formalism
is based on the BGS entropy, it is also desirable  to clarify how the 
the  formalism based on nonextensive entropy may be mapped to the one
based on  BGS entropy. It appears that the choice of Tsallis entropy is not 
appropriate to meet this requirement.

In this paper, we show that the proposed mapping 
from nonextensive to extensive framework (both transformations
(\ref{ne}) and (\ref{lagin})) can be derived
by applying  the generalized zeroth law of thermodynamics for the nonextensive
entropy and the assumption of a Legendre Transform
structure for extensive entropy. Note that the
intensive character of the variables defined by (\ref{lagin})
was pointed out in \cite{abemartinez01} by applying the generalized zeroth law 
of thermodynamics to the Tsallis entropy at equilibrium.
The essential part in the present approach is that the explicit form of 
nonextensive entropy is not assumed a priori, but is derived from
the condition of equilibrium. In this sense, the zeroth law of
thermodynamics fixes the general form of the nonextensive entropy
as a specific function of an extensive entropy. 
Moreover, it will be seen that  the form of extensive constraints 
(whether usual averages or normalised biased averages) decides
that the mapped onto extensive entropy is BGS or Renyi entropy.
  
Consider the thermodynamic equilibrium between two systems $A$ and $B$,
   characterised  by a state of maximum
nonextensive entropy $\tilde{S}^{*}_{q}$  with fixed values of the 
extensive quantities 
\be
E_k(A+B) = E_k(A) + E_k(B), \qquad  k=1,...,n.
\label{extcs}
\ee
As in standard thermodynamics, we define the thermodynamic
equilibrium entropy for the nonextensive case to be
$k_{\rm B}\tilde{S}^{*}_{q}(E_1,...,E_k,..., E_n)$, where
$k_{\rm B}$ is the Boltzmann's constant. Apparently, 
$\tilde{S}^{*}_{q}$ is the 
explicit entropy function in terms of the given extensive
constraints  and is obtained from the equilibrium
distribution $\{p_{i}^{*}\}$ following from the maximisation of
$\tilde{S}_q (p_i)$.  

Now assuming that the joint entropy $\tilde{S}^{*}_{q} (A+B)$ 
satisfies (\ref{nad}), we can
 make the variations $\delta \tilde{S}^{*}_{q} (A+B)$
and $\delta E_k(A+B)$ vanish for the equilibrium state, 
 to obtain 
\be
 [1 + (1-q)\tilde{S}^{*}_{q} (B) ] {\partial \tilde{S}^{*}_{q}(A) \over \partial E_k(A)}
   =  [1 + (1-q)\tilde{S}^{*}_{q} (A) ] {\partial \tilde{S}^{*}_{q}(B) \over \partial E_k(B)},
 \label{vars}
\ee
where $\tilde{S}^{*}_{q}(A)$ implies 
$\tilde{S}^{*}_{q}(E_1(A),...,E_k(A),..., E_n(A))$
and similarly for the system $B$.
On rearranging (\ref{vars}), we can write
\ba
{1 \over [1 + (1-q)\tilde{S}^{*}_{q} (A) ]} {\partial \tilde{S}^{*}_{q}(A)  
\over \partial E_k(A)} &=&
{1 \over [1 + (1-q)\tilde{S}^{*}_{q} (B) ]} {\partial \tilde{S}^{*}_{q}(B) 
\over \partial E_k(B)} \nonumber \\
&=&\eta_k.
\label{rear}
\ea
As each side of the above equation pertains to independent systems $A$ and $B$,
we equate it to a constant $\eta_k$. 
Dropping the index $A$ or $B$, we have the differential equation for nonextensive
entropy of each subsystem
\be
{1 \over [1 + (1-q)\tilde{S}^{*}_{q} ]}{\partial \over \partial E_k}
\tilde{S}^{*}_{q} (E_1,...,E_k,...,E_n)  = \eta_k,
\label{deq}
\ee
which can be integrated to give
\be
 { {\rm ln}\; [1 + (1-q)\tilde{S}^{*}_q] \over (1-q)} = \eta_k E_k + \psi,
\label{int}
\ee
where, the constant of integration $\psi$ is by definition, independent of $E_k$ and
in general, $\psi \equiv \psi(E_1,...,E_{k-1},E_{k+1},...,E_n;\eta_k)$.
Choosing $\psi$ to be a thermodynamic potential equivalent to
a free energy with the property $E_k = {\partial \psi \over \partial \eta_k}$, 
the rhs of (\ref{int}) defines the Legendre transform 
$S^*(E_1,...,E_k,...,E_n)$ of $\psi$ with the
property $\eta_k = {\partial S^* \over \partial E_k}$ \cite{callenbook}.
 In other words,
the quantity on rhs of (\ref{int}) is taken to be an entropy function.
It can be easily verified that lhs of (\ref{int}) defines an extensive
quantity ($S^*$). Thus we obtain the transformation (\ref{ne}). 
Conversely, the explicit form of the maximum nonextensive entropy is  
given by
\be
\tilde{S}^{*}_{q} = {e^{ (1-q)S^* } -1 \over (1-q) }.
\label{tils}
\ee
Using the concavity property of $\tilde{S}^{*}_{q} (\{E_k\})$, we infer that
 $S^*(\{E_k\})$ is concave, provided $q \le 1 $. 

We consider the two types of averaging schemes: \\
(i) {\it usual averages}, implying fixed values of $\sum_i p_i E_{k,i}$
given as constraints. Then the entropy $S^*$  
can be taken as the standard BGS entropy.\\
(ii) {\it normalised biased averages}, as in Tsallis' statistical mechanics.
Then the extensive entropy function may be chosen to be Renyi entropy, which
satisfies the Legendre transform structure with this type of averaging scheme. 
Here we are free to choose Renyi entropy of order $q^{\prime}$. Then the  entropy
obtained from (\ref{tils}) is a two parameter $(q,q^{\prime})$ dependent  
nonextensive entropy, which for $q=q^{\prime}$, reduces to Tsallis entropy.
For concavity, we also require $q^{\prime} \le 1 $.
  
Next we focus on the equilibrium probability distribution obtained from the 
maximisation of  entropy form (\ref{tils}). Below 
we discuss for the case of more general $(q,q^{\prime})$-entropy. First note that
maximising $S^{(R)}_{q^{\prime}}$, the 
Renyi entropy of order $q^{\prime}$, under the
normalised biased scheme with fixed values of
\be  
{\sum_i  p_{i}^{q^{\prime}} E_{k,i} \over \sum_j p_{j}^{q^{\prime}} }
=E_{k}^{(q ^{\prime})}, \qquad (k= 1, ..., n) 
\label{cons2}
\ee
gives the following equilibrium distribution \cite{lenzi00}
\be
 p_i\sim \left[1 - (1-q^{\prime})\sum_{k=1}^{n} 
\eta_{k}^{(\rm R)} (E_{k,i}- E_{k}^{(q ^{\prime})}) \right]^{1/(1-q^{\prime})}.
\label{eqpd2}
\ee
Now if we maximise the nonextensive entropy
\be
\tilde{S}_{q} = {e^{ (1-q)S^{(R)}_{q^{\prime}} } -1 \over (1-q) }.
\label{tils2}
\ee
under the constraints (\ref{cons2}), we get a similar 
distribution as  (\ref{eqpd2}) with the same exponent 
${1/(1-q^{\prime})}$, but with Lagrange multiplier 
$\tilde{\eta}_{k}$ related as $\tilde{\eta}_{k}=
\eta_{k}^{(\rm R)} [1 + (1-q)\tilde{S}^{*}_{q}$], 
where $\tilde{S}^{*}_{q}$ is the maximum
value of (\ref{tils}).
This is evident,  as
the nonextensive entropy (\ref{tils2}) is a continuous, monotonic
increasing function of the extensive entropy $S^{(R)}_{q^{\prime}}$
and so will give the same equilibrium distribution as the latter
under identical constraints.  
However, an important point to note here is that
the exponent of the power-law type equilibrium distribution
is {\it not} necessarily related to the degree of  nonextensivity of
the entropy \cite{johal99pp}, equivalent to $(1-q)$ in the present case.
Finally, for $q^{\prime}=1$, we get case (i) above, 
the constraints reducing to the standard
mean values and extensive entropy is the BGS. 
In this case, the maximisation of $\tilde{S}_{q}$
gives {\it exponential} distributions.

Next, we discuss the mapping between Lagrange 
multipliers of the entropy maximisation problems and 
the intensive variables, in a general  
 context. We assume that for 
a general entropic function ${S}$, the relation  
${\partial {{S}}\over \partial E_k}={\eta}_k $ defines the
Lagrange parameter associated with the constraint quantity
$E_k$, occuring in the maximum entropy problem. 
If the entropy is extensive (additive), then from zeroth law
of thermodynamics, we derive that the Lagrange
multiplier ${\eta}_k$ is also the intensive
variable conjugate to $E_k$.
However, this equivalence breaks down if the entropy
becomes non-extensive. Thus consider a more
general entropy written as $\tilde{S}\equiv F({S})$,
a nonlinear, continuous, monotonic increasing function of ${S}$. 
Then due to the fact that
${\partial \tilde{S}/ \partial E_k} = ({d \tilde{S}/ d {S}})
({\partial {S}/ \partial E_k})$,
we can write as  
\be 
\tilde{\eta}_{k} = {d {\tilde{S}}\over {d S}} {\eta}_k,
\label{lmt} 
\ee
i.e. the Lagrange multipliers of the two maximisation problems 
can be related this way. 
Now if ${S}$ is an extensive entropy, then
(\ref{lmt}) also defines the transformation between the Lagrange
multiplier of a nonextensive entropy and the corresponding intensive
variable, equivalent to (\ref{lagin}).  

Further insight into this relationship can be gained by
using the notion of composability of entropy \cite{hotta99}.
An arbitrary entropic form $\tilde{S}$ is  defined to be composable, if 
the total $\tilde{S}(A,B)$ for the composite system can be written as
$\tilde{S}(A,B) = f[\tilde{S}(A),\tilde{S}(B)]$,
where $f[\cdot ]$ is a certain bivariate function 
of the $C^2$ class and symmetric in its arguments,
$ f[\tilde{S}(A),\tilde{S}(B)] = f[\tilde{S}(B),\tilde{S}(A)]$.
Suppose this general entropy determines the equilibrium
between subsystems $A$ and $B$ under the given extensive
constraints (\ref{cons1}). By equating the variations of the total
entropy and total value of the constraint quantity to zero, 
we get
\be
{\partial f[\tilde{S}(A),\tilde{S}(B)] \over
\partial \tilde{S}(A) } {\partial \tilde{S}(A)
\over \partial E_k(A)} =
{\partial f[\tilde{S}(A),\tilde{S}(B)] \over
\partial \tilde{S}(B) } {\partial \tilde{S}(B)
\over \partial E_k(B)}.  
\label{var0}
\ee
To establish an intensive variable common to the
two subsystems in equilibrium, it is essential 
to have the following factorization \cite{abepre01}
\ba
{\partial f[\tilde{S}(A),\tilde{S}(B)] \over
\partial \tilde{S}(A) } &=& k[\tilde{S}(A),\tilde{S}(B)]\,
 g[\tilde{S}(A)] \,h[\tilde{S}(B)],\label{factgh} \\
{\partial f[\tilde{S}(A),\tilde{S}(B)] \over
\partial \tilde{S}(B) } &=& k[\tilde{S}(A),\tilde{S}(B)]\,
 h[\tilde{S}(A)] \,g[\tilde{S}(B)],
\label{facthg}
\ea
where $g[\cdot ]$ and $h[\cdot ]$ are some functions, in particular
$h[\cdot ]$ will be required to be a differentiable one.
The function $k[\cdot]$ is not in the factorised form.
Then it can be shown that
\be
{1\over g[\tilde{S}(A)]} {d h[\tilde{S}(A)] \over
d \tilde{S}(A)} =
{1\over g[\tilde{S}(B)]} {d h[\tilde{S}(B)] \over
d \tilde{S}(B)} = \omega,
\label{gtoh}
\ee
where $\omega$ is a constant. Moreover,
$k[\tilde{S}(A),\tilde{S}(B)]= G(f[\tilde{S}(A),\tilde{S}(B)])$,
where $G(\cdot)$ is an arbitrary function.  Now the 
identification of an intensive variable is possible only
if it is independent of the function $k[\cdot]$. Therefore,
using (\ref{factgh}), (\ref{facthg}), and
(\ref{gtoh}) in (\ref{var0}), we  must have
\ba
{1\over \omega}  {1\over {h[\tilde{S}(A)]}}{d h[\tilde{S}(A)] \over
d \tilde{S}(A)} {\partial \tilde{S}(A)
\over \partial E_k(A)} &=&
{1\over \omega} {1\over {h[\tilde{S}(B)]}}{d h[\tilde{S}(B)] \over
d \tilde{S}(B)} {\partial \tilde{S}(B)
\over \partial E_k(B)} \nonumber \\
& =& \eta_k,
\label{intr}
\ea
where $\eta_k$ is the intensive variable conjugate
to the constraint $E_k$. The constant $\omega$ has been
kept as in principle, it can vanish. Using the definition of 
Lagrange multiplier of $\tilde{S}$, we obtain
for each subsystem
\be
{1\over \omega} {1\over  {h[\tilde{S}]}}{d h[\tilde{S}] \over
d \tilde{S}} \tilde{\eta}_k = \eta_k.
\label{lmgen}
\ee
This is the most general relation between the
Lagrange multiplier of a composable entropy
and the corresponding intensive variable as 
derived from the zeroth law of thermodynamics.

The function $h$ can be fixed further if we assume that
the general composable entropy $\tilde{S}$
is some monotonic increasing function of 
an extensive entropy $S$. In other words, 
assuming that (\ref{lmt}) holds and using it in (\ref{lmgen}), 
we obtain after integration 
\be
h[S] = {\rm exp}(\omega S +c).
\label{exph}
\ee 
Note that  $h$ comes out as an explicit function 
of the extensive entropy $S$. Lastly, we remark 
for the special case of $k[\cdot]$ as constant function
equal to unity.  
Using (\ref{gtoh}) in (\ref{factgh}) or (\ref{facthg})
and solving for the composability function, we get
\be
f[\tilde{S}(A),\tilde{S}(B)] ={1\over \omega} 
h[S(A)]\,h[S(B)] + c^{\prime}.
\label{fsolv}
\ee
Choosing the constants of integration $c=0$ and $c^{\prime}=-1/\omega$ 
in the above, gives  back the solution (\ref{tils})
for nonextensive entropy with $\omega = (1-q)$.
 
Concluding, it has been observed recently in literature that
thermodynamic structure based on the nonextensive Tsallis entropy can
be mapped to the one based on extensive Renyi entropy. In the 
present paper, we have studied this mapping by establishing the 
zeroth law of thermodynamics, which is shown to determine
 not only the transformation between
Lagrange multipliers and intensive variables,
but also between nonextensive and extensive
entropy. 
The said mapping can be made to the BGS or
Renyi entropy based extensive framework, depending on the
form of extensive constraints. We also remarked that
the exponent in Tsallis-type power-law distributions,
is not necessarily related to the nonextensivity
of entropy. Rather  nonextensivity of entropy seems only  to make  
the Lagrange multipliers entering the maximum entropy problem 
as non-intensive. We have discussed the relation between
Lagrange multiplier and intensive variables for the general
case of composable entropy and determined the form of 
function $h$ when the composable entropy is a monotonic
increasing function of an extensive entropy. 
The present analysis treats 
nonextensivity of entropy keeping the external constraints
as extensive. A more general study can be made by making
the constraints also as nonextensive. The obtained 
relations between for example, the Lagrange multipliers
and intensive variables will also be then generalised.

The financial support from Alexander von Humboldt
Foundation, Bonn, Germany, is gratefully acknowledged. 

\end{document}